# Spin relaxation characteristics in Ag nanowire covered with various oxides

S. Karube[1, 2], H. Idzuchi[1, 2], K. Kondou[2], Y. Fukuma[2, 3], and Y. Otani[1, 2, a)]

[1] *Institute for Solid State Physics, University of Tokyo, Kashiwa 277-8581, Japan*

[2] *Center of Emergent Matter Science, RIKEN, 2-1 Hirosawa, Wako 351-0198, Japan*

[3] *Frontier Research Academy for Young Researchers, Kyushu Institute of Technology, Iizuka 820-8502, Japan*

We have studied spin relaxation characteristics in a Ag nanowire covered with various oxide layers of $Bi_2O_3$, $Al_2O_3$, $HfO_2$, MgO or $AgO_x$ by using non-local spin valve structures. The spin-flip probability, a ratio of momentum relaxation time to spin relaxation time at 10 K, exhibits a gradual increase with an atomic number of the oxide constituent elements, Mg, Al, Ag and Hf. Surprisingly the $Bi_2O_3$ capping was found to increase the probability by an order of magnitude compared with other oxide layers.
This finding suggests the presence of an additional spin relaxation mechanism such as Rashba effect at the $Ag/Bi_2O_3$ interface, which cannot be explained by the simple Elliott-Yafet mechanism via phonon, impurity and surface scatterings. The $Ag/Bi_2O_3$ interface may provide functionality as a spin to charge interconversion layer.

___________________________

a) Electronic mail: yotani@issp.u-tokyo.ac.jp

Spin relaxation in the transport is one of the most important characteristics which determines the performance of spintronic devices such as lateral spin valves (LSVs). The relaxation mechanism in non-magnetic metal (NM) with weak spin-orbit interaction (SOI) was first discussed by Elliott and Yafet.[1,2] According to their theory, the spin relaxation takes place with some probability during momentum scattering events caused by phonons, impurities and surfaces. The spin relaxation have been experimentally examined by means of conduction electron spin resonance, weak localization and non-local spin valve measurements.[3-5] Each contribution to the spin-flip probability in a thin film or a nanowire can be evaluated by the temperature-dependent measurements.

Recently we have reported the suppression of surface spin relaxation in Ag nanowire by MgO capping, of which effect was considered to be caused by the reduction of effective SOI at the Ag/MgO interface.[6] It would be beneficial if one could utilize this interface SOI similarly to the recent experiment by Rojas Sánchez *et al*. demonstrating that the interfacial spin to charge conversion appears at the Bi(111)/Ag interface via the inverse Rashba-Edelstein effect.[7] This finding of the interfacial spin to charge conversion is expected to bring about new methods of manipulating magnetization or detecting spin currents in spintronic devices, and opened a question on the spin transport properties of the system which has rarely been studied so far.[7,8] In this research phase, therefore, one of the important steps is to evaluate the potential of interfacial spin transport by testing what kind of materials are appropriate. Here we show the spin-flip probabilities $\varepsilon$ in various Ag/Oxide hetero-structures determined from the analysis of non-local spin valve (NLSV) signals.

Figure 1(a) shows a typical SEM image of a fabricated LSV device which consists of two ferromagnetic $Ni_{80}Fe_{20}$ (Py) injector and detector nanowires bridged by a Ag nanowire with Py/MgO/Ag junctions. The Ag wire was covered with an oxide layer of $Bi_2O_3$, $Al_2O_3$, or $HfO_2$. The fabrication procedure is as follows: Firstly a $Si/SiO_2$ (300 nm) substrate was spin-coated with a 500 nm-thick methyl-methacrylate (MMA) and a 50 nm-thick poly-methyl-methacrylate (PMMA). Then the device structure was patterned by e-beam lithography. In the present study, shadow evaporation was employed to obtain clean interfaces in Py/MgO/Ag junctions. A 20 nm-thick Py and a few nm thick MgO were deposited at the angle of 45 degrees from the substrate by means of e-beam evaporation. Next a 50 nm-thick Ag wire was deposited at the angle of 90 degrees from the substrate cooled by liquid $N_2$ to protect the MMA/PMMA resist from the radiation heating from the evaporation source. The Py/MgO and Ag wires were deposited in separate chambers (base pressure: ~$10^{-7}$ Pa) to prevent magnetic impurities mixed in the Ag wire during deposition. Finally, a 5 nm-thick $Bi_2O_3$, $Al_2O_3$, or $HfO_2$ was deposited on the Ag wire.

For determining the spin diffusion length in Ag, various devices with different separation distances between two Py wires varied from 300 nm to 1500 nm were fabricated. The widths of Py/MgO and Ag wire are 150 nm and 120 nm, respectively.

Non-local spin injection measurements were performed on the LSVs with Py/MgO/Ag junctions to evaluate the spin diffusion length in the Ag nanowire covered with the oxide layer. A dc current of 100 μA was applied to Py/MgO/Ag junction to generate spin accumulation in the Ag wire. The external magnetic field was applied in the range from -1100 to +1100 Oe along the Py wire. Figure 1(b) shows the typical non-local spin signals in the LSVs with the Ag wire covered with Al$_2$O$_3$ or Bi$_2$O$_3$ capping. The separation distance between two Py wires $L$ is 300 nm and the measurement temperature $T$ is 10 K. The spin signals $\Delta R_S$ for LSVs with Bi$_2$O$_3$ and Al$_2$O$_3$ are 1.5 mΩ and 18 mΩ, respectively. The magnitude of the spin signal for the Bi$_2$O$_3$ capping is ten times smaller than that for the Al$_2$O$_3$ capping. This result suggests the presence of strong spin relaxation due to the Ag/Bi$_2$O$_3$ interface.

Figure 2 shows the separation distance dependence of the spin signals in the Ag wires covered with Al$_2$O$_3$ or Bi$_2$O$_3$. The fitting line for the Bi$_2$O$_3$ capping is much steeper than that for the Al$_2$O$_3$ capping. We also estimated the spin diffusion length by using Eq. (1),[9,10]

$$\Delta R_S = \frac{4 R_S^{Ag} \left[ \frac{P_I}{1-P_I^2} \left( \frac{R_I}{R_S^{Ag}} \right) + \frac{P_{Py}}{1-P_{Py}^2} \left( \frac{R_S^{Py}}{R_S^{Ag}} \right) \right]^2 e^{-L/\lambda_{Ag}}}{\left[ 1 + \frac{2}{1-P_I^2} \left( \frac{R_I}{R_S^{Ag}} \right) + \frac{2}{1-P_{Py}^2} \left( \frac{R_S^{Py}}{R_S^{Ag}} \right) \right]^2 - e^{-2L/\lambda_{Ag}}} \quad (1)$$

where $P_I$ and $P_{Py}$ are the spin polarizations of the MgO interface and ferromagnetic metal (FM), respectively. $R_S^{Ag} = \rho_{Ag} \lambda_{Ag}/t_{Ag} w_{Ag}$, $R_S^{Py} = \rho_{Py} \lambda_{Py}/w_{Ag} w_{Py}/(1-P_{Py}^2)$ are the spin resistances for NM and FM, respectively. $R_I$ is the MgO interface resistance, where $\rho_i, \lambda_i, t_i,$ and $w_i (i = $ Ag or Py) are the resistivity, the spin diffusion length, the thickness of the wire, and the width of the wire, respectively. Results of the spin signal in Fig. 2 were analyzed by using Eq. (1). The values of $\rho_{Py}$ = 35 μΩcm and $R_I$= 1 and 0.1 Ω for Bi$_2$O$_3$ and Al$_2$O$_3$ were respectively determined by using the LSVs fabricated and the reported values of $\lambda_{Py} = 5$ nm, $P_{Py} = 0.35$ were used in the fitting procedure.[11,12] We then obtained $\lambda_{Ag}^{Ag/Bi_2O_3} = 127 \pm 13$ nm, $\lambda_{Ag}^{Ag/Al_2O_3} = 450 \pm 44$ nm, $P_I^{Ag/Bi_2O_3} = 0.23 \pm 0.04$, $P_I^{Ag/Al_2O_3} = 0.20 \pm 0.01$. We found that the spin diffusion length of the Ag wire with the Bi$_2$O$_3$ capping drastically decreased.

In order to take into account the influence of the quality of the Ag wire on the spin relaxation process, we estimated the spin-flip probability $\varepsilon = \tau_e/\tau_{\text{sf}}$, where $\tau_e = m_e/ne^2\rho_{\text{Ag}}$ and $\tau_{\text{sf}} = \lambda_{\text{Ag}}^2/D_{\text{Ag}}$ are the momentum relaxation time and the spin relaxation time with $m_e, n, e, D_{\text{Ag}} = \left(e^2 N(\varepsilon_{\text{F}})\rho_{\text{Ag}}\right)^{-1}$ the electron mass, the electron density, the elementary charge and the diffusion constant, respectively; we used the density of state in Ag at the Fermi energy, $N(\varepsilon_{\text{F}}) = 1.55 \times 10^{22}/\text{eV}/\text{cm}^3$.[13] Table I summarizes characteristic spin transport properties for different oxide capping layers obtained from our analyses. The values of $\varepsilon$ for the Ag with MgO capping and non-capping are reported in our previous work.[6] For the Ag wire without capping, we assume that the oxide layer of AgO$_x$ is naturally formed on the surface.[14] The spin relaxation in non-magnetic metals has been explained by Elliott-Yafet mechanism.[1,2] In this mechanism, $\varepsilon$ can be given in the relation, $1/\tau_{\text{sf}}(= \varepsilon/\tau_e) = 1/\tau_{\text{sf}}^{\text{ph}} + 1/\tau_{\text{sf}}^{\text{imp}} = \varepsilon_{\text{ph}}/\tau_e^{\text{ph}} + \varepsilon_{\text{imp}}/\tau_e^{\text{imp}}$, where the notations, "ph" and "imp", respectively mean phonon and impurity including grain boundary and surface scatterings. The spin-flip probability of the Ag wire covered with Bi$_2$O$_3$ is ten times higher than the other cases. The momentum relaxation time for the Bi$_2$O$_3$ capping is shorter than the case for other capping layers and therefore one might suspect that Bi impurities are mixed into the Ag wire and contributes to the spin relaxation. In this regard, we confirmed that no Bi impurities contribute to the spin relaxation by an observation of spin to charge conversion in Py/Ag/Bi$_2$O$_3$ trilayer system using spin pumping method and the sign of the conversion efficiency, i.e. spin Hall angle (SHA) was positive.[15] If Bi diffuses into Ag and forms AgBi diluted alloy, a sign of SHA due to extrinsic spin Hall effect should be negative.[16] Therefore, we believe that the Bi diffusion is not a dominant origin of the spin relaxation in the Ag wire.

We here discuss the interface scattering contribution $\varepsilon_{\text{imp}}^{\text{surf}}$ in the probability $\varepsilon_{\text{imp}}$. The magnitude of $\varepsilon_{\text{imp}}$ can be directly deduced from the experimental data at low temperatures where the phonon contribution to the scattering is negligible.[17] As shown in the table, the mean free path of electrons in the Ag wire is larger or comparable to the thickness of ~50 nm.[6] The transport properties are thus considered to be affected mostly by their NM/oxide interfaces. Remarkable is that the $\varepsilon_{\text{imp}}$ of the Ag wire with Bi$_2$O$_3$ capping $(= 47.2 \times 10^{-3})$ is an order of magnitude larger than that with Al$_2$O$_3$ $(= 5.41 \times 10^{-3})$, which is also significantly larger than the reported values of $\varepsilon_{\text{imp}}$ ranging from 0.2 to $4.0 \times 10^{-3}$ for Ag wires and thin films tabulated in Table 3 of Ref. 18. The spin relaxation at 10 K can be described as a sum of two main contributions from capping-dependent surfaces and capping-independent impurities such as grain boundaries and dislocations. This leads to the relation: $1/\tau_{sf}^{imp} = \varepsilon_{imp}/\tau_e^{imp} = \varepsilon_{imp}^{grain}/\tau_e^{grain} + \varepsilon_{imp}^{surf}/\tau_e^{surf}$, where $\tau_e^{grain}$ and $\tau_e^{surf}$ are respectively the momentum relaxation times due to the impurities and the surfaces. The spin-flip probability is then given by

$$\varepsilon_{imp} = \left(\frac{\tau_e^{imp}}{\tau_e^{grain}}\right)\varepsilon_{imp}^{grain} + \left(\frac{\tau_e^{imp}}{\tau_e^{surf}}\right)\varepsilon_{imp}^{surf}$$

$$= \left(\frac{\tau_e^{surf}}{\tau_e^{grain} + \tau_e^{surf}}\right)\varepsilon_{imp}^{grain} + \left(\frac{\tau_e^{grain}}{\tau_e^{grain} + \tau_e^{surf}}\right)\varepsilon_{imp}^{surf} \quad (2)$$

At $T = 10$ K, the electrons behave as ballistic particles in the Ag nanowire due to larger or comparable mean free path of the electrons to the thickness, alternatively get many opportunities to collide with the surface. Therefore $\tau_e^{surf}$ is considered much shorter than $\tau_e^{grain}$, thus the spin-flip probability $\varepsilon_{imp}$ can be approximated as $\varepsilon_{imp}^{surf}$. Now in order to compare the surface spin relaxation properties of different capping oxide layers, we plot the spin-flip probability of the surface contribution $\varepsilon_{surf}(= \varepsilon_{imp} - \varepsilon_{imp}^{MgO})$ for different capping layers as shown in Fig. 3. Here, the $\varepsilon_{imp}^{MgO}$ is the spin-flip probability obtained from a Ag nanowire covered with MgO layer,[6] which mainly includes the contribution of the grain boundary $\varepsilon_{imp}^{grain}$ because the surface contribution is suppressed. In Al$_2$O$_3$, AgO$_X$, HfO$_2$ cases, the values of $\varepsilon_{surf}$ are in the same order of magnitude, implying that the surface spin relaxation is also explained by the Elliott-Yafet mechanism. Moreover the spin-flip probability $\varepsilon_{surf}$ increases monotonically with increasing an atomic number $Z$ except for Bi$_2$O$_3$ case. Here, remarkable is that the $\varepsilon_{surf}$ of the Ag wire with Bi$_2$O$_3$ capping is an order of magnitude larger than the other oxide case.

Lastly we discuss the possible mechanism of additional spin relaxation in the Ag/Bi$_2$O$_3$ interface. Large Rashba splitting has recently been observed in Ag(111)/Bi surfaces.[19,20] This interfacial Rashba effect is originated from a partial charge density near Bi atomic nuclei generating large effective electric fields.[20] In our case, e-beam deposited poly-crystalline Ag layers may have preferred (111) orientation because of the lowest surface energy.[21] In addition, for Ag/Bi$_2$O$_3$ interface, O$^{2-}$ ions may enhance the partial charge density due to its high electronegativity.

In the point of view of the spin relaxation, at the interface where the Rashba effect exists, the spin relaxation time was found to be the same order of the momentum relaxation time because of the spin-momentum locking,[7,22] while in conventional metals the spin relaxation time (~ps) is much longer than the momentum relaxation time (~fs). We thus believe that the spin relaxation in the case of Ag/Bi$_2$O$_3$ could be more strongly affected than the cases with other oxide capping layers by spin-momentum locking originated from the interfacial Rashba effect.

In summary, we have studied the spin relaxation in Ag nanowires with various oxide capping layers by means of non-local spin injection method. We experimentally found that the spin-flip probabilities $\varepsilon$ in the Ag wire covered with various oxides except for Bi$_2$O$_3$ at low temperature are of the same order of magnitude, and gradual increase with an atomic number of the oxide constituent elements. Most importantly we observed a large spin-flip probability in the Ag wire with Bi$_2$O$_3$ capping (Ag/Bi$_2$O$_3$ interface). This fact implies the presence of the additional interfacial spin relaxation mechanism caused by the interfacial Rashba effect. This may provide a novel metal/insulator interface where the interconversion between charge and spin takes place.

**Acknowledgments**
This work was supported by Grant-in-Aid for Scientific Research on Innovative Area, "Nano Spin Conversion Science" (Grant No. 26103002) and Japan Society for the Promotion of Science through Program for Leading Graduate Schools (MERIT).

Table I. Characteristic parameters of the spin relaxation in Ag with various oxide capping layers, where the values for the cases of MgO and AgO$_X$ capping layers are taken from previous work.[6] $Z$ of M, $\rho_{Ag}$, $\tau_e$, $\lambda_e$, $\lambda_{Ag}$, $\tau_{sf}$, and $\varepsilon$ mean the atomic number of the metal element in the oxide, the resistivity, the momentum relaxation time, the mean free path of electrons, the spin diffusion length, the spin relaxation time, and the spin-flip probability, respectively. The sample of HfO$_2$ capping was annealed at 500 °C for 30 minutes in N$_2$ (97 %) and H$_2$ (3 %) atmosphere. The samples of Bi$_2$O$_3$ and Al$_2$O$_3$ capping were as deposited.

| MO$_X$ | $Z$ of M | $\rho_{Ag}$ (μΩcm) | $\tau_e$ (fs) | $\lambda_e$ (nm) | $\lambda_{Ag}$ (nm) | $\tau_{sf}$ (ps) | $\varepsilon_{imp}$ ($= \tau_e/\tau_{sf} \times 10^{-3}$) |
|---|---|---|---|---|---|---|---|
| MgO [6] | 12 | 0.90 | 68.0 | 94.5 | 851 | 16.2 | 4.20 |
| Al$_2$O$_3$ | 13 | 1.50 | 40.8 | 56.7 | 450 | 7.50 | 5.41 |
| AgO$_X$ [6] | 47 | 1.00 | 61.2 | 85.1 | 667 | 11.0 | 5.54 |
| HfO$_2$ | 72 | 0.79 | 77.5 | 108 | 692 | 9.40 | 8.25 |
| Bi$_2$O$_3$ | 83 | 1.80 | 34.0 | 47.3 | 127 | 0.70 | 47.2 |

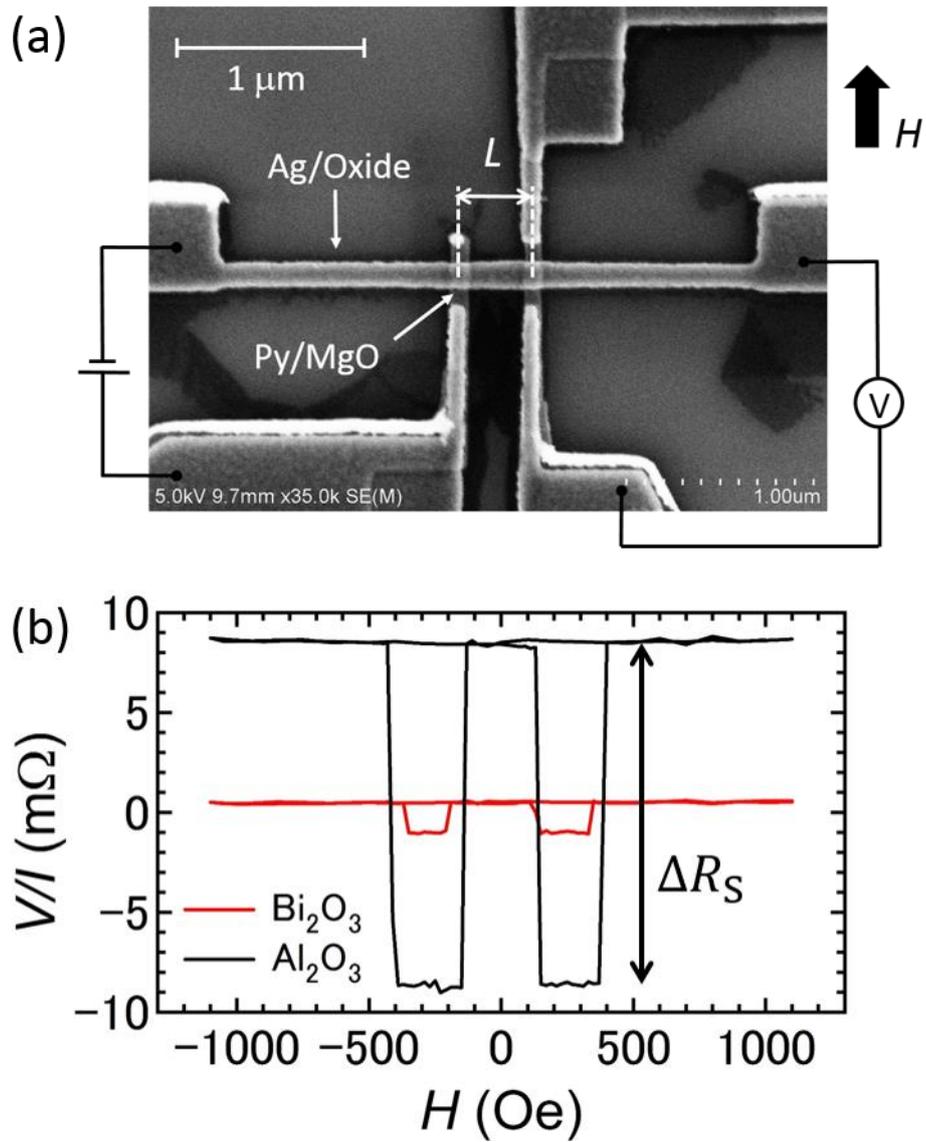

Figure 1. (a) SEM image of a prepared LSV with a schematic circuit for non-local measurement. (b) Typical non-local spin valve signals at $T = 10$ K as a function of external magnetic field for LSVs with Ag wires covered with $Al_2O_3$ (black line) and $Bi_2O_3$ capping (red line), with $L = 300$ nm.

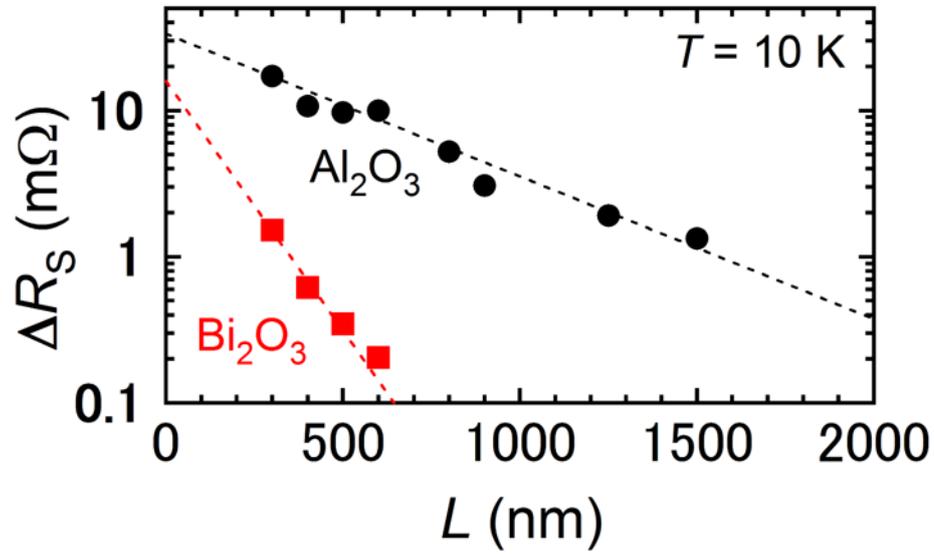

Figure 2. Non-local spin signals $\Delta R_S$ as a function of separation $L$ between two Py electrodes. Black circles (red squares) correspond to Ag wire with $Al_2O_3$ ($Bi_2O_3$) capping. Dashed lines are fitting lines to data by using Eq. (1).

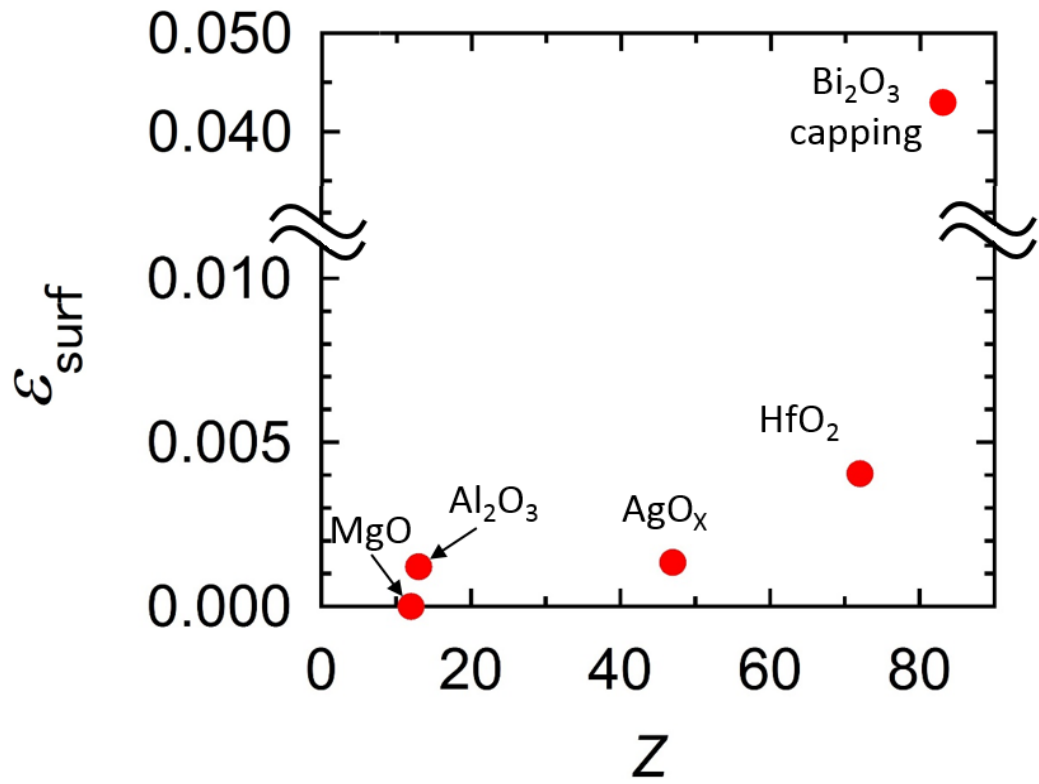

Figure 3. Surface spin-flip probability $\varepsilon_{\text{surf}}(=\varepsilon_{\text{imp}} - \varepsilon_{\text{imp}}^{\text{MgO}})$ as a function of $Z$ for various oxide capping layers where $Z$ is an atomic number of the metal element in the oxide, and the label in the graph means each capping case.